\documentclass{elsart}

\usepackage{amsmath,amssymb,bm,graphicx}

\begin{document}

\begin{frontmatter}



\title{Evidence for an alpha cluster condensed state
in ${}^{16}{\rm O}(\alpha,\alpha')$ at 400 MeV}


\author[kyushu]{T.~Wakasa}
\ead{wakasa@phys.kyushu-u.ac.jp}
\author[kyushu]{E.~Ihara}
\author[rcnp]{K.~Fujita}
\author[riken]{Y.~Funaki}
\author[rcnp]{K.~Hatanaka}
\author[rcnp]{H.~Horiuchi}
\author[cyric]{M.~Itoh}
\author[jaeri]{J.~Kamiya}
\author[rostock]{G.~R\"opke}
\author[miyazaki]{H.~Sakaguchi}
\author[rcnp]{N.~Sakamoto}
\author[rcnp]{Y.~Sakemi}
\author[ipn]{P.~Schuck}
\author[rcnp]{Y.~Shimizu}
\author[yukawa,riken]{M.~Takashina}
\author[riken]{S.~Terashima}
\author[rcnp]{A.~Tohsaki}
\author[tit]{M.~Uchida}
\author[kyushu2]{H.~P.~Yoshida}
\author[rcnp]{M.~Yosoi}

\address[kyushu]{Department of Physics, Kyushu University,
Fukuoka 812-8581, Japan}
\address[rcnp]{Research Center for Nuclear Physics, Osaka University,
Osaka 567-0047, Japan}
\address[riken]{The Institute of Physical and Chemical Research,
Wako, Saitama 351-0198, Japan}
\address[cyric]{Cyclotron and Radioisotope Center, Tohoku University,
Sendai, Miyagi 980-8578, Japan}
\address[jaeri]{Accelerator Group, Japan Atomic Energy Research Institute,
Tokai, Ibaraki 319-1195, Japan}
\address[rostock]{Institut f\"ur Physik, Universit\"at Restock,
D-18051 Rostock, Germany}
\address[miyazaki]{Department of Engineering, Miyazaki University,
Miyazaki 889-2192, Japan}
\address[ipn]{Institut de Physique, Nucl\'eaire, F-91406 Orsay Cedex,
France}
\address[yukawa]{Yukawa Institute for Theoretical Physics,
Kyoto University,
Kyoto 606-8502, Japan}
\address[tit]{Department of Physics, Tokyo Institute of Technology,
Tokyo 152-8550, Japan}
\address[kyushu2]{
Center for Research and Advancement in Higher Education,
Kyushu University,
Fukuoka 810-8560, Japan}

\begin{abstract}
 Inelastic $\alpha$ scattering on ${}^{16}{\rm O}$ is 
studied at 400 MeV by using an ice target.
 Near the $4\alpha$ breakup threshold of 14.4 MeV, 
a broad peak is observed at an
excitation energy of $13.6\pm 0.2$ MeV 
with a width of $0.6\pm 0.2$ MeV.
 The spin-parity is estimated to be $0^+$ from the momentum-transfer 
dependence.
 The observed width is significantly larger than those of the 
neighboring $0^+$ states indicating a state with a well-developed $\alpha$
cluster structure.
 The magnitude of the cross section is sensitive to the 
density distribution of the constituent $\alpha$ clusters.
 The observed cross section is consistent with the theoretical 
prediction for the $\alpha$ cluster condensed state characterized 
by its dilute density distribution 
with a large root-mean-square radius of about 4.3 fm.
\end{abstract}

\begin{keyword}
$(\alpha,\alpha')$ scattering \sep
$\alpha$ cluster condensation \sep
\PACS 25.55.Ci \sep 21.60.Gx \sep 27.20.+n
\end{keyword}
\end{frontmatter}

 In the last decade, Green's function Monte Carlo (GFMC) calculations 
have successfully reproduced ground and low-lying excited 
state energies of light nuclei up to $A$ = 10 
\cite{prc_62_014001_2000,prc_66_044310_2002,npa_751_516c_2005}
with an accuracy of 1--2\% using realistic two-nucleon 
\cite{prc_51_38_1995}
and 
three-nucleon potentials \cite{prl_74_4396_1995,prc_64_014001_2001}.
 One of the most impressive results of these so-called 
first-principles calculations for $A$ = 8 is that 
the intrinsic density distribution for the ground state of 
${}^{8}{\rm Be}$ has two peaks representing a $2\alpha$
cluster structure \cite{prc_62_014001_2000}.
 These two clusters are separated by about 4 fm, and thus
they form a dilute system.
 It is natural to expect that there are also 
loosely bound dilute states with $3\alpha$ and $4\alpha$ clusters
as excited states in ${}^{12}{\rm C}$ and ${}^{16}{\rm O}$, 
respectively.
 For ${}^{12}{\rm C}$, several $3\alpha$ cluster calculations 
\cite{ns_fukushima,ptp_uegaki,npa_351_456_1981}
almost a quarter century ago 
have shown that the second $0^+$ state has a larger root-mean-square (rms) 
radius than 
the ground state by about 1 fm.
 These results imply that this $0^+$ state has a gas-like 
structure of $3\alpha$ clusters predominantly in 
relative $S$ waves \cite{ptp_51_1266_1974,ptp_53_447_1975}.

 Recently, using a new $\alpha$ cluster wave function, 
a $4\alpha$ cluster state of dilute density has been 
theoretically predicted in ${}^{16}{\rm O}$ near the
$4\alpha$ breakup threshold \cite{prl_87_192501_2001}.
 The same wave function has also been used to predict
a 3$\alpha$-cluster state with dilute density 
in ${}^{12}{\rm C}$ near the $3\alpha$ breakup threshold
\cite{prl_87_192501_2001,prc_67_051306_2003},
and there are strong indications \cite{epja_28_259_2006} 
that the second $0^+$ state is this state.
 This new $\alpha$ cluster wave function can represent a 
condensation of $\alpha$ clusters in a spherically symmetric 
state.
 Because the $\alpha$ cluster is a boson,
an excited state with dilute density composed of 
weakly interacting $\alpha$ clusters in relative $S$
waves can be considered as a $\alpha$ cluster condensed 
(ACC) state.
 This idea is based on theoretical investigations on the 
possibility of $\alpha$ particle condensation in low-density 
nuclear matter \cite{prl_80_3177_1998}.
 It should be noted that, 
in the case of finite self-conjugate $4n$ nuclei, 
coherent phenomena such
as super-fluidity observed in a Bose--Einstein condensed state 
are not expected to occur since $n$ is very small, e.g., $n=3$ in ${}^{12}{\rm C}$.
 Nevertheless the idea of $\alpha$ cluster condensation
in nuclei is very important since the ACC 
state does seem to exist in self-conjugate 
$n\alpha$ nuclei \cite{prl_87_192501_2001,prc_69_024309_2004}.
 Therefore, it is very important to search for and investigate the 
theoretically predicted $4\alpha$ cluster condensed state 
in ${}^{16}{\rm O}$.
 In this Letter we report experimental evidence
for a broad $0^+$ state at an excitation energy of 
$E_x$ = $13.6\pm 0.1$ MeV with a width of 
$\Gamma$ = $0.6\pm 0.2$ MeV which can be interpreted as 
being the predicted exotic ACC state.

 Measurements were carried out using the 
West--South Beam Line (WS-BL) \cite{nim_a482_79_2002} 
and the Grand Raiden (GR) spectrometer \cite{nim_a422_484_1999}
at the Research Center for Nuclear Physics, Osaka University.
 The WS-BL provides a double-achromatic beam with 
zero lateral and angular dispersions.
 The beam energy was $T_\alpha$ = 400 MeV (100 MeV/A) and 
its energy spread was 108 keV.
 The beam bombarded a windowless and self-supporting 
ice (${\rm H_2O}$) target \cite{nim_a459_171_2001} 
with a thickness of 9 ${\rm mg/cm^2}$.
 The thickness was determined by comparing the elastic 
yield from the ${\rm SiO_2}$ target with a thickness 
of 2.0 ${\rm mg/cm^2}$.
 Scattered $\alpha$ particles from the target were momentum analyzed 
by the high-resolution GR spectrometer 
with a typical resolution of about 160 keV FWHM.
 The yields of the scattered $\alpha$ particles were extracted using the 
peak-shape fitting program {\sc allfit} \cite{allfit}.

\begin{figure}
\begin{center}
\includegraphics[width=0.9\linewidth,clip]{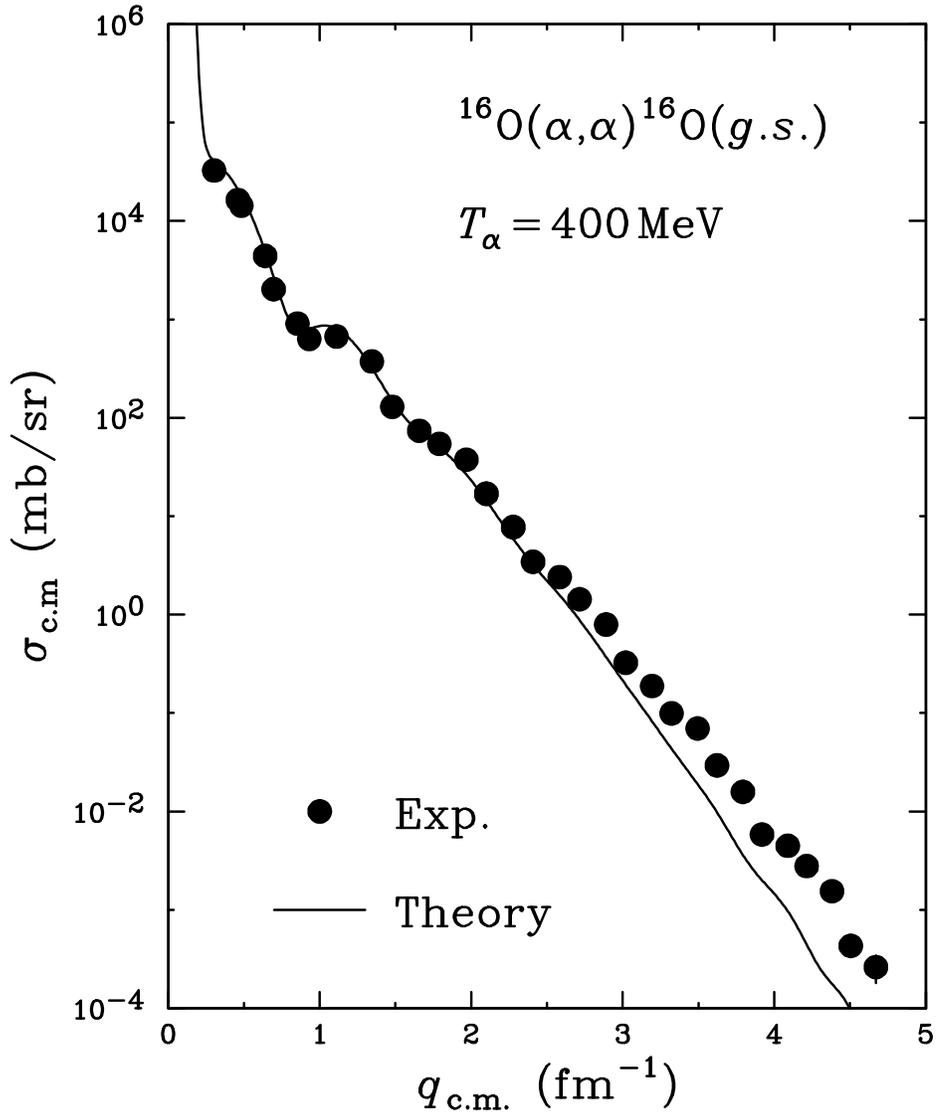}
\end{center}
\caption{
 Measurement of the cross section (solid circles) 
for ${}^{16}{\rm O}(\alpha,\alpha)$ at $T_\alpha$ = 400 MeV.
 The solid curve is the theoretical prediction using the 
single folding potential for ${}^{16}{\rm O}$, 
as explained in the text.
\label{fig:elastic}}
\end{figure}

 The elastic differential cross sections of ${}^{16}{\rm O}$ are 
shown in Fig.~\ref{fig:elastic}.
 The data decrease almost monotonically as a function of 
momentum transfer over a momentum-transfer region up to 4.7 ${\rm fm^{-1}}$.
 The data were analyzed by a phenomenological 
optical model potential (OMP).
 We adopt the single-folding model based on the 
nucleon--$\alpha$ interaction, in which an empirical 
nucleon--$\alpha$ potential \cite{prc_61_034312_2000} 
is folded with the nucleon density of ${}^{16}{\rm O}$.
 Note that the resulting OMP is a complex potential 
because the nucleon--$\alpha$ potential is complex.
 The solid curve in Fig.~\ref{fig:elastic} shows the result 
using the OMP generated by the folding.
 The single-folding potential reproduces the data well
up to $q_{\rm c.m.}$ $\approx$ 3 ${\rm fm^{-1}}$ and 
slightly underestimates the data for 
$q_{\rm c.m.}$ $\gtrsim$ 3 ${\rm fm^{-1}}$.
 This underestimation is believed to be due to the energy dependence 
of the nucleon--$\alpha$ interaction and could be 
resolved by adjusting the parameters of the 
single-folding potential.
 However, in the following, we use the OMP for the analysis 
of the ACC without modification because the ACC data have been 
measured only up to $q_{\rm c.m.}$ $\approx$ 2.3 ${\rm fm^{-1}}$ where the 
elastic data are well reproduced by this OMP.

\begin{figure}
\begin{center}
\includegraphics[width=0.9\linewidth,clip]{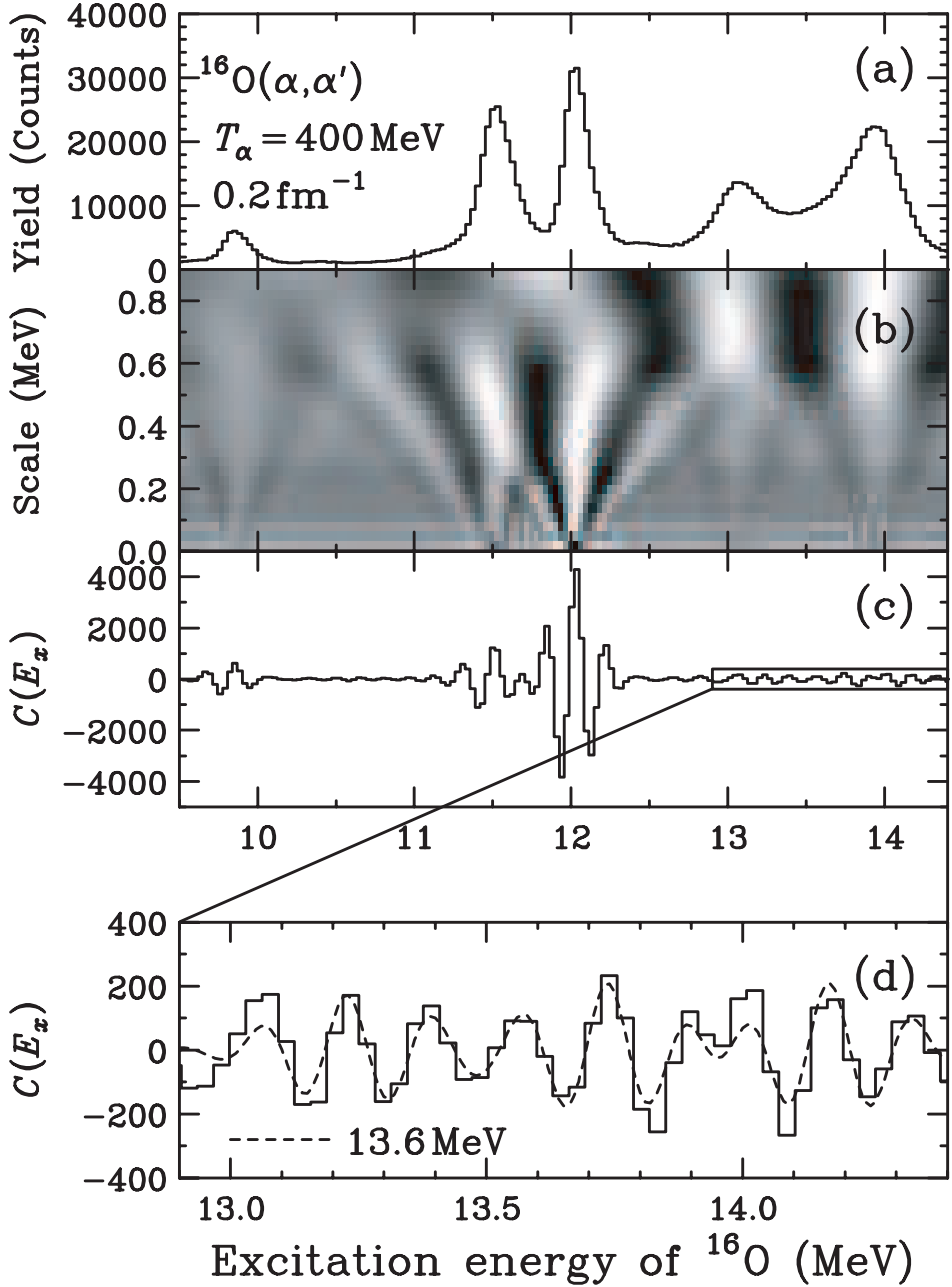}
\end{center}
\caption{
 (a) Spectrum of ${}^{16}{\rm O}(\alpha,\alpha')$ scattering
at $T_\alpha$ = 400 MeV and $q_{\rm c.m.}$ = 0.2 ${\rm fm^{-1}}$.
 (b) Result of CWT analysis as a function 
of both excitation energy and scale parameter.
 (c) Wavelet coefficients at a scale parameter of 0.12 MeV.
 (d) Magnification of the region in (c) around the new state at 13.6 MeV.
\label{fig:wavelet}}
\end{figure}

 Figure~\ref{fig:wavelet}(a) shows the excitation energy 
spectrum of ${}^{16}{\rm O}(\alpha,\alpha')$ scattering 
including $\theta_{\rm c.m.}$ = $0^{\circ}$.
 The effects of the finite solid angle of GR have been 
taken into account in all the momentum transfer 
$q_{\rm c.m.}$ values reported here. 
For Fig.~\ref{fig:wavelet}(a), $q_{\rm c.m.}$ $\simeq$ $0.2\,\mathrm{fm}^{-1}$ is obtained.
 At this small $q_{\rm c.m.}$, 
the $J^{\pi}$ = $0^+$ state at $E_x$ = 12.0 MeV is the most prominent peak
and 
the $J^{\pi}$ = $2^+$ state at $E_x$ = 11.5 MeV also forms 
a prominent peak.
 Both broad peaks at $E_x$ $\simeq$ 13.0 and 14.0 MeV consist 
of several peaks. An ACC state is expected to be in this 
region.
 We performed peak fitting using {\sc allfit} 
to find evidence of the ACC state.
 In the peak fitting, the narrow peaks of ${}^{16}{\rm O}$ were described by a 
standard hyper-Gaussian line shape and the peaks with intrinsic 
widths were described as Lorentzian shapes convoluted with a 
resolution function based on the narrow peaks.
 The positions and widths were taken from Ref.~\cite{nudat2}.
 It was found that the region below $E_x$ $\simeq$ 13 MeV is 
well reproduced in the peak fitting, whereas the 
region around $E_x$ $\simeq$ 13.6 MeV is significantly 
under-predicted because there is no known state in this region
\cite{nudat2}.
 The under-prediction is common at all momentum transfers and 
indicates the existence of a new state in this region.

 A wavelet analysis technique was employed to investigate the structure of the new state.
 This technique has been developed in signal theory 
\cite{wavelets} 
and has also been used in nuclear physics for the extraction 
of characteristic scales in the fine structure of giant resonances
\cite{prl_93_122501_2004,prl_96_012502_2006}.
 The excitation energy spectrum $\sigma(E)$ was folded with a 
chosen wavelet function $\Psi$ and the coefficients
\begin{equation}
C(E_x,\Delta E)=\frac{1}{\sqrt{\Delta E}}\int\sigma(E)
\Psi\left(\frac{E-E_x}{\Delta E}\right)dE
\label{eq:c}
\end{equation}
were obtained in the continuum wavelet transform (CWT), 
where $E_x$ is the excitation energy and $\Delta E$
is the scale parameter of $\Psi$.
 We used the Morlet type function consisting of a cosine function with 
a Gaussian envelope,
\begin{equation}
\Psi(x) = \frac{1}{\sqrt[4]{\pi}}\cos(kx)
\exp\left(-\frac{x^2}{2}\right)\ ,
\quad
k=\frac{5}{2\pi}
\end{equation}
which is often used for the wavelet function 
\cite{prl_93_122501_2004,prl_96_012502_2006}.

 Figures~\ref{fig:wavelet}(b)--(d) show the results of CWT for the data 
at $q_{\rm c.m.}$ = 0.2 ${\rm fm^{-1}}$.
 The correlation of the wavelet coefficients 
$C$ in Eq.~(\ref{eq:c}) is displayed in Fig.~\ref{fig:wavelet}(b)
as functions of $E_x$ and $\Delta E$.
 For the bumps at $E_x$ $\simeq$ 13.0 and 14.0 MeV consisting 
of several peaks, two maxima at scales of about 
0.1 and 0.7 MeV are observed.
 The larger value corresponds to the total width of the bumps 
and the smaller value reflects the width of the 
constituent peaks as determined by the 
experimental energy resolution.
 These results indicate that the CWT technique can be applied 
to the investigation of the fine structure of the bump 
excited by $(\alpha,\alpha')$ scattering.
 Figure~\ref{fig:wavelet}(c) shows $C$ at a scale of 
0.12 MeV as a function of excitation energy.
 The structures of the Morlet type function are clearly found 
at $E_x$ $\approx$ 9.8, 11.5, and 12.0 MeV, which correspond to 
the known states with $J^{\pi}$ = $2^+$, $2^+$, and $0^+$, 
respectively.
 An uneven structure of $C$ is identified at $E_x$ $\gtrsim$ 
13 MeV, displayed magnified in Fig.~\ref{fig:wavelet}(d).
 The structure at $E_x$ $\simeq$ 13.6 MeV indicates the existence of the
new state because there is no known state in this region.
 The observed structure is well reproduced by assuming the 
excitation energy of the new state to be $E_x$ = 13.6 MeV, as shown 
by the dashed curve in Fig.~\ref{fig:wavelet}(d).
 The reproducibility is sensitive to the excitation energy 
of the new state and thus the uncertainty is estimated
to be $\pm 0.1$ MeV.
 It is rather difficult to 
determine the width of the new state from the results of CWT because of interference from the neighboring bumps at 
$E_x$ $\simeq$ 13.0 and 14.0 MeV.

\begin{figure}
\begin{center}
\includegraphics[width=0.9\linewidth,clip]{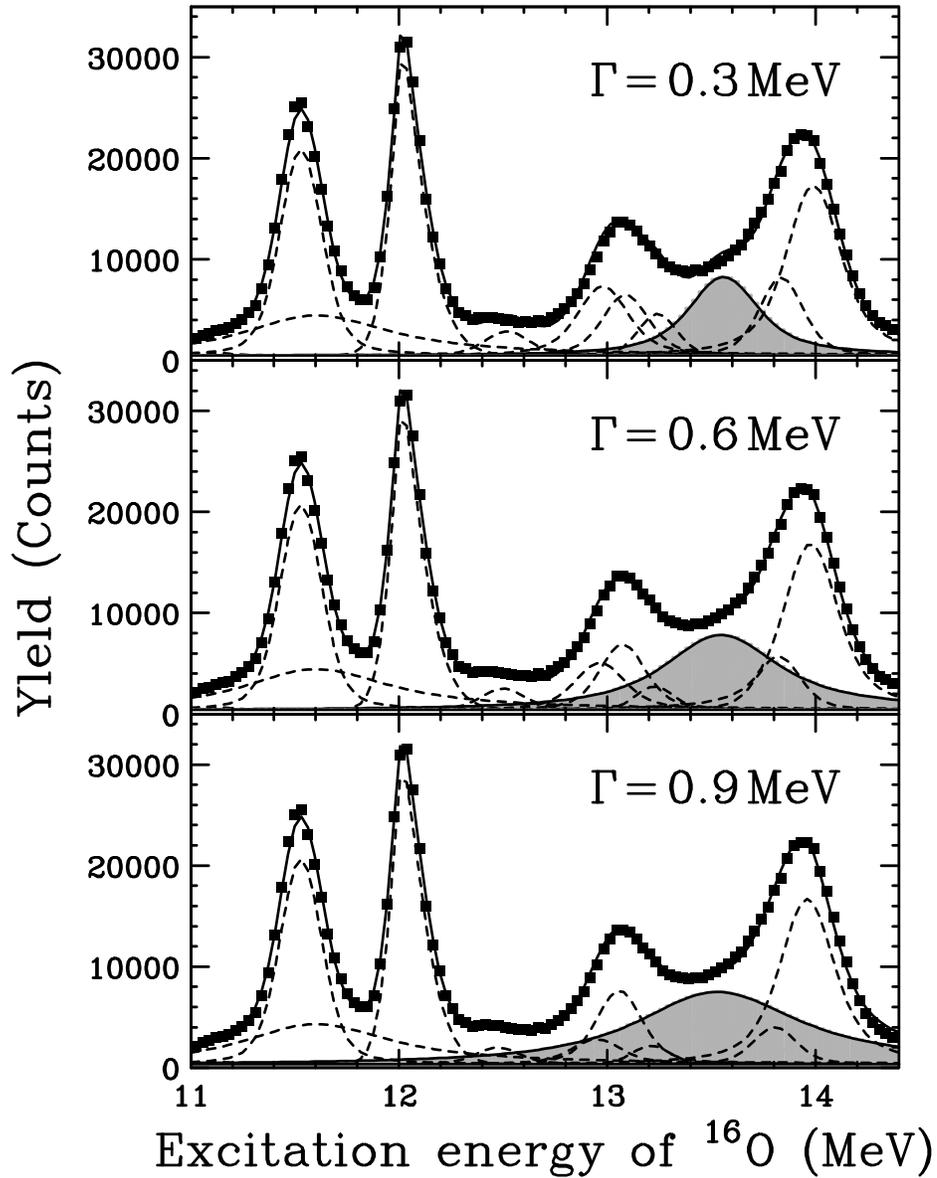}
\end{center}
\caption{
 Results of peak fitting of ${}^{16}{\rm O}(\alpha,\alpha')$ 
at $T_\alpha$ = 400 MeV and $q_{\rm c.m.}$ = 0.2 ${\rm fm^{-1}}$
with hyper-Gaussian and Lorentzian peaks and a continuum,
 with the width $\Gamma$ of the new state at 13.6 MeV taken to be 
(a) 0.3 MeV, (b) 0.6 MeV, and (c) 0.9 MeV.
\label{fig:width}}
\end{figure}

\begin{figure}
\begin{center}
\includegraphics[width=0.9\linewidth,clip]{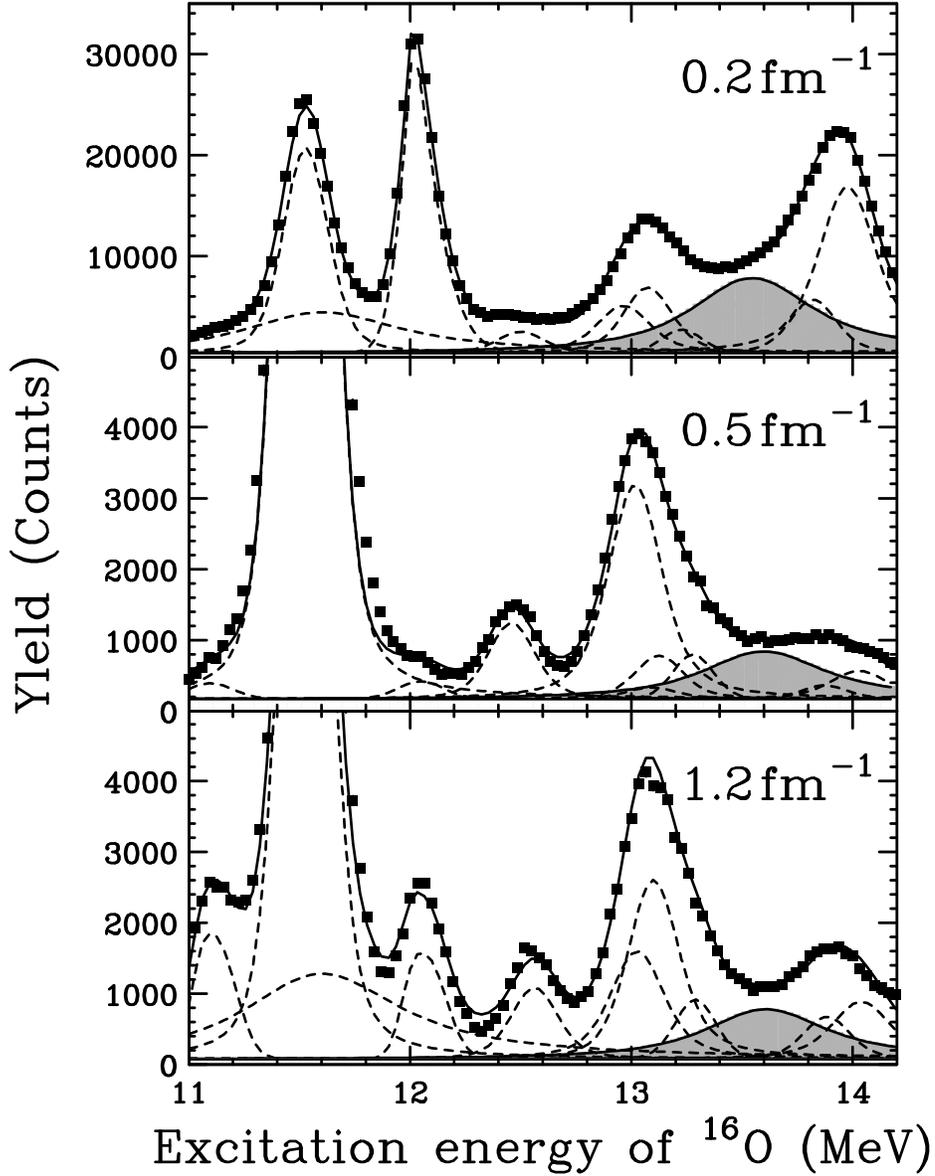}
\end{center}
\caption{
 Reproduction of the excitation energy spectra for 
${}^{16}{\rm O}(\alpha,\alpha')$ 
at $T_\alpha$ = 400 MeV and 
a momentum transfer $q_{\rm c.m.}$ of 
(a) 0.2 ${\rm fm^{-1}}$,
(b) 0.5 ${\rm fm^{-1}}$, and
(c) 1.2 ${\rm fm^{-1}}$.
 The width of the new state at 13.6 MeV is taken to be as 0.6 MeV.
\label{fig:spec}}
\end{figure}

 In order to deduce the width of the new state, 
we performed peak fitting in this region 
by adding the new state at $E_x$ = 13.6 MeV, 
and determined its width so as to reproduce the experimental data.
 Figure~\ref{fig:width} shows the results of the fitting 
at $q_{\rm c.m.}$ = 0.2 ${\rm fm^{-1}}$ 
for $\Gamma$ = 0.3, 0.6, and 0.9 MeV.
 The dashed curves represent the fits to the individual 
peaks while the straight line and solid curve 
represent the background and the sum of the peak 
fitting, respectively.
 In the case of $\Gamma$ = 0.3 MeV, the region around the new 
state could not be well reproduced, and in the case of 
$\Gamma$ = 0.9 MeV, the region at $E_x$ $\simeq$ 14.4 MeV 
is overestimated.
 The best reproduction is achieved for $\Gamma$ = 0.6 MeV 
and its uncertainty is estimated to be about 0.2 MeV, calculated 
considering this result and the uncertainty of the excitation 
energy.
 We also performed peak fitting for other 
momentum-transfer data; the results are displayed 
in Fig.~\ref{fig:spec} for 
$q_{\rm c.m.}$ $\simeq$ 0.2, 0.5, and 1.2 ${\rm fm^{-1}}$.
 At all momentum transfers, reasonable
reproduction of the data is achieved by including the new state at 
$E_x$ = $13.6\pm 0.1$ MeV with $\Gamma$ = $0.6\pm 0.2$ MeV.

 It should be noted 
that the width $\Gamma$ = $0.6\pm 0.2$ MeV of the new state is 
significantly larger than that of the neighboring $0^+$ states
at $E_x$ = 12.0 and 14.0 MeV, with $\Gamma$ = 1.5 and 185 keV 
\cite{nudat2}, respectively.
 A recent theoretical investigation with the new $\alpha$ 
cluster wave function predicts the ACC state of ${}^{16}{\rm O}$ 
to be $E_x$ $\approx$ 14.1 MeV and $\Gamma$ $\approx$ 1.5--1.9 MeV
\cite{mpla_funaki}.
 Quantitative comparisons between experimental and theoretical results
are rather difficult due to the parameter dependence of the 
theoretical predictions.
 However, the calculations show qualitatively 
that the ACC state should 
be around the $4\alpha$ break-up threshold of 14.4 MeV 
with a significantly large $\Gamma$ compared with those 
of the $0^+$ states described in the shell model.
 The ACC state is characterized by the diluteness of the 
constituent $\alpha$ clusters, and this diluteness 
results in the large $\Gamma$ value of the ACC.
 The observed new state is consistent with this theoretical 
prediction for the ACC.
 Note that the transition form factor to the ACC is almost 
independent of the parameters \cite{mpla_funaki}.
 Thus, in the following, a qualitative comparison is made 
of the cross section in order to 
confirm that the new state is the ACC.

\begin{figure}
\begin{center}
\includegraphics[width=0.9\linewidth,clip]{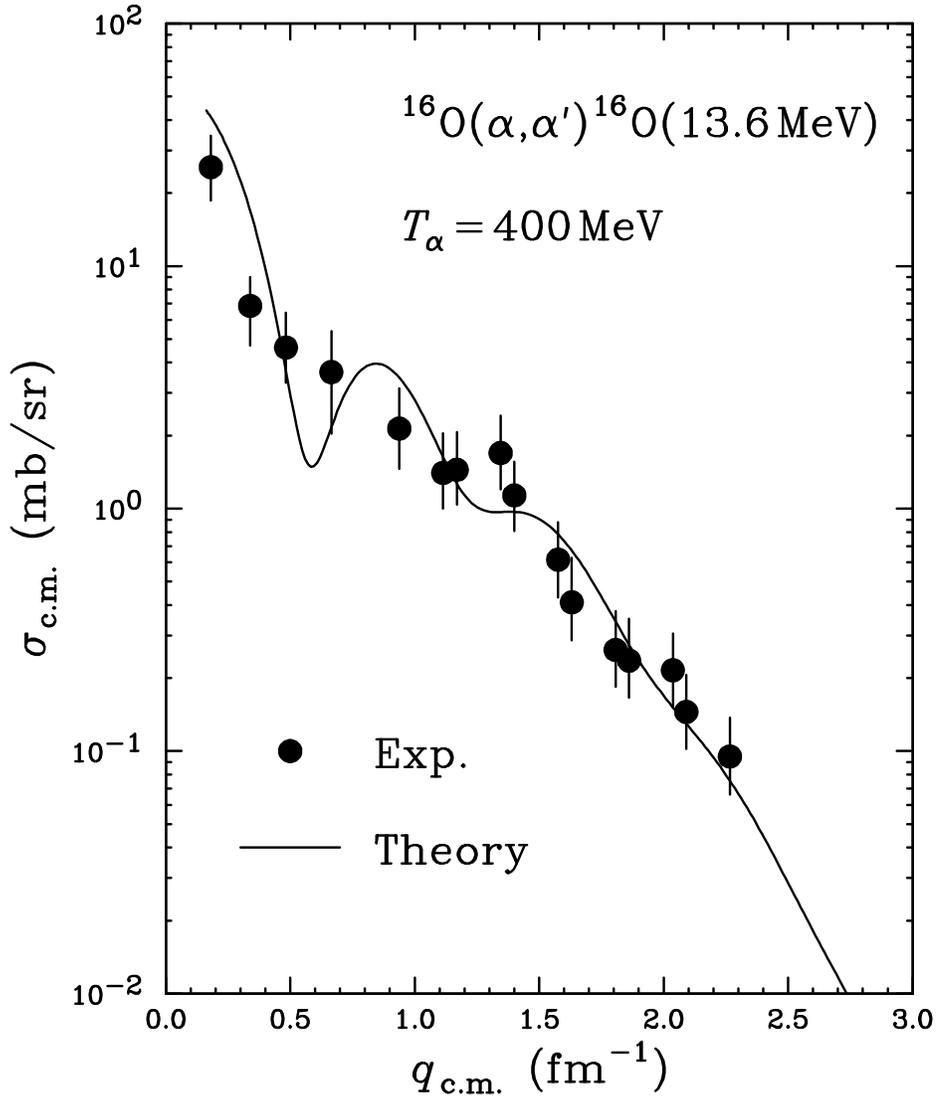}
\end{center}
\caption{
 Measurement of the cross section of 
${}^{16}{\rm O}(\alpha,\alpha'){}^{16}{\rm O}(13.6\,{\rm MeV})$ 
at $T_\alpha$ = 400 MeV.
 The error bars include both statistical and systematic 
uncertainties of the data.
 The solid curve is the theoretical prediction for 
the ACC state in ${}^{16}{\rm O}$,
as described in the text.
\label{fig:acc}}
\end{figure}

 Figure~\ref{fig:acc} shows the cross section of the new
state at $E_x$ = 13.6 MeV as a function of momentum transfer.
 The error bars include both statistical and systematic uncertainties 
of the data, with the systematic uncertainties being dominant.
 The systematic uncertainties are estimated by taking into account 
the uncertainties of both the excitation energy and the 
width of the new state.
 The data cover a momentum-transfer region up to about 2.3 ${\rm fm^{-1}}$, 
and have a maximum at 
$q_{\rm c.m.}$ = 0.2 ${\rm fm^{-1}}$
($\theta_{\rm c.m.}$ $\approx$ $0^{\circ}$). 
 This forward-peaking of the cross section indicates 
that the transferred angular momentum $L$ is 0 and the 
spin parity of the new state is $0^+$ by considering 
the spin-scalar and isospin-scalar nature of $(\alpha,\alpha')$.
 We compare our cross section with a microscopic coupled-channel (MCC)
calculation for the ACC state.
 The OMP is the same as that used for the analysis of the elastic 
scattering data, and the transition form factor is obtained 
using the ACC wave function described in Ref.~\cite{mpla_funaki}.
 The solid curve in Fig.~\ref{fig:acc} is the result of the calculation 
and reproduces the experimental data well without normalization.
 It should be noted that the magnitude of the form factor is 
sensitive to the rms radius of the ACC state, whereas 
its radial dependence is insensitive to the rms radius
\cite{mpla_funaki}.
 Therefore, for the corresponding cross section in $(\alpha,\alpha')$,
the rms radius should be investigated through its magnitude rather than its momentum-transfer 
dependence \cite{prc_takashina}.
 Since we have used an OMP that reproduces the elastic 
scattering data well, we can compare the absolute values 
of both the experimental and theoretical results.
 The good theoretical reproduction without any normalization shows that 
the rms radius of the new state is consistent with that of 
the ACC state used in the calculation.
 Therefore, the rms radius could be estimated to be same as the 
theoretical value of about 4.3 fm \cite{mpla_funaki},
which is significantly larger than that of the ground state 
of ${}^{16}{\rm O}$ of 2.7 fm.
 This large rms radius is due to the diluteness 
of the $\alpha$ clusters, and thus the new state can be 
interpreted as being the ACC state characterized by its 
dilute density distribution.

 In conclusion, we have searched for evidence of the ACC state 
in ${}^{16}{\rm O}$ in ${}^{16}{\rm O}(\alpha,\alpha')$ 
scattering at 400 MeV using an ice target.
 A broad peak was found at $E_x$ = $13.6\pm 0.1$ MeV 
with $\Gamma=0.6\pm 0.2$ and its spin-parity 
$J^{\pi}$ was estimated to be $0^+$ from the 
momentum-transfer dependence.
 The excitation energy is near the $4\alpha$ breakup threshold 
of 14.4 MeV, whereas the width is significantly larger than those 
of neighboring $0^+$ states.
 This large $\Gamma$ indicates that the $\alpha$ cluster structure 
is well-developed in the new state.
 The magnitude of the cross section is sensitive to the diluteness
of the $\alpha$ clusters.
 The observed cross section, the excitation energy, and 
the width all strongly indicate the existence of an ACC state in 
${}^{16}{\rm O}$, characterized by a large rms radius of about 4.3 fm compared with 
that of the ground state of 2.7 fm.

\ack
 The authors gratefully acknowledge the dedicated efforts 
of the RCNP cyclotron crew for providing a good quality beam.
 This work was supported in part by the Grants-in-Aid for
Scientific Research Nos.~14702005 and 16654064
of the Ministry of Education, Culture, Sports, 
Science, and Technology of Japan.

\clearpage

\clearpage

\clearpage

\clearpage

\clearpage

\clearpage

\bibliographystyle{elsart-num}
\bibliography{wakasa}

\end{document}